\documentclass[twocolumn,showpacs,amsmath,amssymb,prb]{revtex4}
\usepackage{graphicx}
\usepackage{bm}
\newcommand{\beq}{\begin{equation}}
\newcommand{\eeq}{\end{equation}}

\newcommand{\bfk}{{\boldsymbol{k}}}
\newcommand{\bfQ}{{\boldsymbol{Q}}}
\newcommand{\bfq}{{\boldsymbol{q}}}
\newcommand{\bfe}{{\boldsymbol{e}}}
\newcommand{\eps}{\varepsilon}

\newcommand{\bpm}{\begin{pmatrix}}
\newcommand{\epm}{\end{pmatrix}}
\renewcommand{\th}{\mathop{{\rm th}}\nolimits}
\newcommand{\sign}{\mathop{{\rm sign}}\nolimits}
\renewcommand{\Re}{\mathop{{\rm Re}}\nolimits}
\renewcommand{\Im}{\mathop{{\rm Im}}\nolimits}

\begin{document}

\title{The phase diagram and the structure of CDW state in high magnetic field in quasi-1D materials:
mean-field approach}
\date{\today }
\author{P. D. Grigoriev}
\affiliation{ National High Magnetic Field laboratory, Florida State
University, Tallahassee, Florida }
 \altaffiliation[Permanent address: ]{L. D. Landau Institute for Theoretical Physics,
 Chernogolovka, Russia}

 \email{grigorev@magnet.fsu.edu}
\author{D. S. Lyubshin}
\affiliation{ L. D. Landau Institute for Theoretical Physics, 142432
Chernogolovka, Russia } \date{\today }

\begin{abstract}

We develop the mean-field theory of a charge-density wave (CDW)
state in magnetic field and study the properties of this state below
the transition temperature. We show that the CDW state with shifted
wave vector in high magnetic field (CDW$_x$ phase) has at least
double harmonic modulation on the most part of the phase diagram. In
the perfect nesting case the single harmonic CDW state with shifted
wave vector exists only in a very narrow region near the tricritical
point where the fluctuations are very strong. We show that the
transition from CDW$_0$ to CDW$_x$ state below the critical
temperature is accompanied by a jump of the CDW order parameter and
of the wave vector rather than by their continuous increase. This
implies a first order transition between these CDW states and
explains the strong hysteresis accompanying this transition in many
experiments. We examine how the phase diagram changes in the case of
imperfect nesting.
\end{abstract}



\maketitle

\section{Introduction}

The properties of metals with a charge-density-wave (CDW) ground
state attract great attention since the fifties (see, e.g.,
monographs [\onlinecite{Gruner,Monceau}]). In quasi-1D metals, the
Fermi surface (FS) consists of two slightly warped sheets separated
by $2k_F$ and roughly possesses the nesting property that leads to
the Peierls instability and favors the formation of the charge- or
spin-density-wave (SDW) state at low temperature.

The mean-field description is known to be unable to describe
strictly 1D conductors, where the non-perturbative methods and
exactly solvable models are of great importance.\cite{1Dtheory}
However, in most materials manifesting CDW or SDW phenomena
 the nonzero value of the electron transfer integral
between conducting chains and the 3D character of the
electron-electron interactions and lattice elasticity reduce the
deviations from the mean-field solution and also make most of the
methods and exactly solvable models developed for strictly 1D case
inapplicable.\cite{Comment1D} The effect of fluctuations in Q1D
metals and their influence on the mean-field description of the
Peierls transition in these metals has been considered in a number
of papers (see e.g. [\onlinecite{HorGutWeger,McKenzieFluct}] and
references therein). It was shown that the interchain dispersion of
electrons strongly damps the fluctuation and validate the mean-field
description. In Q1D organic metals,\cite{Ishi} at which our present
study is mainly aimed, the free electron dispersion near the Fermi
level is, approximately, given by
 \begin{equation}
   \label{dispersion1}
   \eps_\sigma (\bfk ) = \hbar v_F(|k_x|-k_F) -t_{\perp }( \bfk_{\perp })- \sigma H,
 \end{equation}
where $v_F$ and $k_F$ are the Fermi velocity and Fermi momentum in
the chain ($x$) direction, $H\equiv \mu_B B$ is the Zeeman energy,
$\mu_B$ is the Bohr magneton and $B$ is the external magnetic field.
The perpendicular-to-chain term, $t_{\perp }( \bf{k}_{\perp })$, is
much greater than the energy scale of the CDW(SDW) transition
temperature, $T_{c0}$. Only the ''imperfect nesting'' part, $t'_b$,
of $t_{\perp }( \bf{k}_{\perp })$ is of the order of $T_{c0}$ (see
Eqs. (\ref{dispersion}), (\ref{2})). Hence, the criterion for the
mean-field theory to be applicable,\cite{HorGutWeger,McKenzieFluct}
$t_{\perp }\gg T_{c0}$, is reliably satisfied in most Q1D organic
metals.

The mean-field description of the CDW properties is, in many
aspects, very similar to the BCS theory. The pairing of two
electrons in superconductors is replaced in CDW by the pairing of an
electron with the hole on the opposite sheet of the Fermi surface.
The charge and spin coupling constants in CDW (see interaction
Hamiltonian (\ref{Hint})) are analogous to coupling constants in
spin-singlet and spin-triplet channels in superconductivity. The CDW
phase with shifted nesting vector is similar to the non-uniform
superconducting phase (LOFF phase).\cite{FFLOLO,FFLOFF} However,
there are several important differences between these two
many-particle effects. The first difference is that the formation of
a gap in the electron spectrum in CDW leads to an insulating rather
than superconducting state. This happens due to the pinning of the
CDW condensate by crystal imperfections. Hence, the CDW state does
not reveal a superfluid current.\cite{Gruner,Monceau} Other
differences appear, e.g., in the excitation spectrum. In particular,
the lowest energy excitations in magnetic field or at imperfect
nesting are not always the electron pairs as in BCS theory but may
be the "soliton kinks".\cite{Su,Braz,BGS,BGL,BuzTug}

The theoretical investigation of the CDW/SDW properties at a
mean-field level comprises two main branches. The first focuses on
the transition line from the metallic states to CDW/SDW state using
susceptibility calculations. This allows to include many additional
factors into the theoretical model, such as different free electron
dispersion relations, spin and orbital effects of the external
magnetic field, applied pressure etc. It helped to discover and to
explain many beautiful effects such as the field-induced spin- and
charge-density waves (FISDW),\cite{gor84,Mont,FISDWEx,LebFICDW} the
increase of the transition temperature due to the
one-dimensionization of the electron spectrum in high magnetic
field,\cite{gor84,cha96} the CDW state with shifted nesting vector
\cite{Mont,ZBM}) etc. However, all these results cannot be continued
below the transition temperature since they are based on the
electron susceptibility calculations using metallic-state electron
Green's functions. In particular, the calculation of
[\onlinecite{ZBM}] predicts an appearance of the CDW$_x$ phase with
shifted wave vector as the magnetic field exceeds some critical
value. This calculation gives the metal-CDW$_x$ transition line,
$T_c(H)$, and the dependence of the optimal shift $q_x(H)$ of CDW
wave vector on magnetic field at this transition line. However, this
calculation does not extend below the transition temperature, and,
hence, cannot be used to describe the properties and the phase
diagram inside the CDW phase. For example, it neither describes the
structure of the CDW$_x$ phase below $T_c(H)$ nor gives the
CDW$_0$-CDW$_x$ transition line $H_{c1}(T)$ or the kind of this
transition. The mean-field study of CDW in Ref.
[\onlinecite{DietrichFulde}] is applicable only in weak field when
the CDW wave vector does not shift from its zero-field value.

The second branch of investigation involves the soliton physics. It
appeared first to describe the ground state of polyacetylene
\cite{Su,Braz} and then developed into a rather large activity (see
Refs. [\onlinecite{SuReview,BrazKirovaReview}] for a review). In
particular, the soliton structure and the energy spectrum of CDW
state in external magnetic field were considered \cite{BDK}.
However, all these results were only derived at zero temperature and
perfect nesting and are not applicable at temperatures of the order
of transition temperature, $T\sim T_c$. The finite-temperature phase
transition to soliton phase has been considered in Ref.
[\onlinecite{BuzTug}]. However, that phase diagram refers to zero
field and finite shift of the chemical potential from the value
$\mu_0$ corresponding to the commensurate CDW.\cite{BuzTug} The
analysis of the CDW phase diagram in magnetic field has been
performed in the case of perfect nesting and at the electron density
close to half-filling.\cite{Machida} However, this approach is also
unable to describe the incommensurate case far from half-filling. In
particular, it suggests the one-harmonic modulation of charge
density, which is usually unstable in high magnetic field (see
below).

Further theoretical description of the CDW state in magnetic field
became important last years because of the intensive experimental
study of the CDW state in strongly anisotropic organic metals.
\cite{KartsLaukin,Biskup,Christ,Qualls,Andres,HarrCDW,OrbQuant,UncCDW1,UncCDW2,Graf,Graf1}

In the present paper we study the CDW phase diagram and the
properties of the CDW state at finite temperature below the
transition point in the mean-field approximation. This study links
the results of susceptibility calculations and the results from the
soliton approach. We take into account the spin effect of the
external magnetic field and consider the CDW phase with shifted
nesting vector. The ''antinesting'' term in the electron dispersion
is also taken into account. We show that the CDW$_x$ state with
shifted wave vector, proposed in Ref. [\onlinecite{ZBM}], has double
harmonic modulation almost everywhere in the phase diagram (see Fig.
\ref{PhDTri},\ref{PhDTot}). At perfect nesting, $t'_b=0$, the single
harmonic CDW$_x$ state exists only in a very narrow region near the
tricritical point. We show that the transition from CDW$_0$ to
CDW$_x$ state below the critical temperature is accompanied by a
jump of the CDW wave vector rather than by its continuous increase.
This implies a first order transition between the CDW states and
explains the strong hysteresis observed at the kink
transition.\cite{Christ,Qualls,OrbQuant,UncCDW2}

Besides the Zeeman splitting, external magnetic field affects the
orbital electron motion. First, an external magnetic field
perpendicular to conducting layers leads to one-dimensionization of
the electron spectrum in quasi-1D metals \cite{cha96} that improves
the nesting property of the Fermi surface. Second, due to the strong
scattering by the nesting vector, electrons in quasi-1D metal in
magnetic field may form close orbits in momentum space \cite{Mont}.
These orbital effects increase the CDW or SDW transition temperature
and may lead to the field-induced spin-density waves \cite{gor84}
(for a review, see [\onlinecite{Ishi}]). A similar effect for the
charge density waves was also proposed \cite{LebFICDW}. The
interplay between spin and orbital effects of the magnetic field is
quite interesting. It leads, for example, to a series of phase
transitions between the CDW states with quantized nesting
vector.\cite{OrbQuant}

In our present study we disregard the orbital effects of magnetic
field. This limitation, however, is not very restrictive. If the
magnetic field is parallel to the conducting layers, it produces
only the Zeeman splitting and no orbital effect. The orbital
quantization effects are, usually, more subtle than the spin effect.
As an illustration, the quantized nesting phases can be observed
only at very low temperature \cite{OrbQuant} because the effects of
nesting quantization are strongly damped at $\hbar\omega_c =e\hbar
H/m^* c \ll 2\pi T, 2\pi\hbar /\tau ,\Delta $. This damping is
somewhat similar to that of the magnetic quantum oscillations. The
one-dimensionalization of electron spectrum in magnetic field can be
effectively taken into account in the present study by introducing
the magnetic field dependence of the imperfect nesting transfer
integral, $t'(H)$, in the electron dispersion relation
(\ref{dispersion}). A situation which is mathematically equivalent
to the Zeeman splitting of magnetic field without its orbital effect
arises when there are two slightly different Q1D chain system in the
same compound with coupling between the chains of different type.
The quantities $U_c+U_s$ and $U_c-U_s$ play the role of the electron
coupling constant inside the chains and between the chains of
different type, respectively. Such a system occurs in many compounds
as TTF-type organic metals (see, e.g., the discussion on page 196 in
[\onlinecite{BrazKirovaReview}]). Two slightly different types of
chains occur also in $\alpha
$-(Per)$_2$M(mnt)$_2$.\cite{Per1,Graf,Graf1} In the systems with
molecular chains of two types, however, the difference between
optimal CDW wave vectors on different chains is fixed by the
molecular structure and does not vary as in the case of external
magnetic field which gradually separates the Fermi surfaces with two
different spin components.

\section{The Model and the mean-field theory of CDW in magnetic field.}

\subsection{The model}

We consider a quasi-1D metal with dispersion (\ref{dispersion1}) and
$t_{\perp }( \bfk_{\perp })$ given by the tight-binding model:
\begin{equation}
   \label{dispersion}
   t_{\perp }( \bfk_{\perp }) = -2t_b\cos (k_y b)
     -2t'_b\cos (2k_y b)-2t_c\cos (k_z c_z),
 \end{equation}
where $b$ and $c_z$ are the lattice constants in $y$- and
$z$-directions respectively. The dispersion along the $z$-axis is
assumed to be much weaker than the dispersion along $y$-direction.
Therefore, we omit the second harmonic $\propto \cos (2k_z c_z)$ in
the dispersion relation (\ref{dispersion}). Since the terms
$2t_c\cos (k_z c_z)$ and $2t_b\cos (k_y b)$ do not violate the
perfect nesting condition
 \begin{equation}
   \label{2}
   \varepsilon (\boldsymbol{p}+\boldsymbol{Q})=-\varepsilon
   (\boldsymbol{p}),
 \end{equation}
they do not influence the physics discussed below unless the nesting
vector become shifted in $y$-$z$ plane. We do not consider such a
shift in the present study.

The electron Hamiltonian is
$$
  \hat H = \hat H_0 + \hat H_{\rm int},
$$
with the free-electron part
 \begin{equation}
   \label{H0}
   \hat H_0 = \sum_{\bfk\sigma} \eps_\sigma(\bfk) a^\dag_\sigma(\bfk)
   a_\sigma(\bfk)
 \end{equation}
and the interaction part
 \begin{eqnarray}
   \label{Hint}
   \hat H_{\rm int} = \frac{1}{2}
   \sum _{\bfk\bfk'\bfQ\sigma\sigma'} V_{\sigma \, \sigma '}(\bfQ )
   a^\dag_\sigma(\bfk+\bfQ)a_\sigma(\bfk)\nonumber \\
   \times a^\dag_{\sigma'}(\bfk'-\bfQ) a_{\sigma'}(\bfk').
 \end{eqnarray}
This Hamiltonian does not include the orbital effect of magnetic
field on quasi-1D electron spectrum which may be important in the
case of substantially imperfect nesting and strong magnetic field
perpendicular to the easy-conducting plane (see introduction). Later
on we will be interested mainly in the interaction at the wave
vector $\bfQ$ close to the so-called nesting vector $\bfQ_0$, which
is usually chosen as
 \begin{equation}
   \label{Q0}
\boldsymbol{Q}_0=(\pm 2k_F,\pi /b, \pi /c).
 \end{equation}
The deviations $\bfQ-\bfQ_0$ that will be considered below are of
the order of $\max \{ H,t'_b\} /\hbar v_F \ll k_F$, so that for such
small deviations the interaction function
 \begin{equation}
   \label{vQ}
V_{\sigma \, \sigma '}(\bfQ )\approx V_{\sigma \, \sigma '}(\bfQ
=\bfQ_0)= U_c-U_s\sigma\sigma'
\end{equation}
The coupling constants $U_c$ and $U_s$ are the same as in Ref.
[\onlinecite{ZBM}]. Subscripts $c$ and $s$ distinguish charge and
spin coupling constants. The Hamiltonian (\ref{Hint}) can be
obtained from the extended Hubbard model when the e-e scattering by
the momenta close to the nesting wave vector $\bfQ_0$ is only taken
in to account.

\subsection{Mean field approach}

First, we note that the terms with $\bfQ=0$ in (\ref{Hint}) only
renormalizes the chemical potential and will not be omitted in
subsequent calculations (all sums over $\bfQ $ do not include $\bfQ
=0$. We introduce the thermodynamic Green's function
 \begin{equation}
   \label{Gkk}
   g_\sigma(\bfk',\bfk,\tau-\tau') =
     \langle T_\tau\{
       a^\dag_\sigma(\bfk',\tau') a_\sigma(\bfk, \tau)
     \}\rangle,
 \end{equation}
where the operators are taken in the Heisenberg representation, and
the average
 \begin{equation}
   \label{DQs}
   D_{\bfQ\sigma} = \sum_{\bfk} g_\sigma(\bfk-\bfQ,\bfk,-0)
   - \delta(\bfQ, 0) n_\sigma.
 \end{equation}
After defining
 \begin{equation}
  \label{DeltaQs}
  \Delta_{\bfQ\sigma} = \sum_{\sigma'}(U_c - U_s\sigma\sigma')
    D_{\bfQ\sigma'}
 \end{equation}
one obtains in the mean-field approximation
 \begin{equation}
  \label{H}
  \hat H_{\rm int}=\sum_{\bfQ\bfk\sigma} a^\dag_{\sigma}(\bfk+\bfQ) a_{\sigma}(\bfk)
    \Delta_{\bfQ\sigma}
  - \frac{1}{2}\sum_{\bfQ\sigma} D_{-\bfQ\sigma} \Delta_{\bfQ\sigma}.
 \end{equation}
Hermiticy of the Hamiltonian requires
$\Delta_{-\bfQ\sigma}=\Delta^*_{\bfQ\sigma}$.  It is now
straightforward to write down the equations of motion which in the
frequency representation take the form
 \begin{eqnarray}
   \label{GEw}
   [i\omega-\eps_\sigma(\bfk)] g_\sigma(\bfk',\bfk,\omega)
   -\sum_\bfQ \Delta_{\bfQ\sigma} g_\sigma(\bfk',\bfk-\bfQ,\omega)
   \nonumber \\
   =\delta(\bfk',\bfk).\ \ \
 \end{eqnarray}

\subsection{The Cosine Phase}

Now we consider the solution with $\Delta_{\bfk\sigma} \neq 0$ only
for $\bfk = \pm \bfQ$, where $\bfQ = 2k_F\bfe_x+ (\pi/b) \bfe_y
+\bfq $ with $|\bfq |\ll k_F$.  If we neglect the scattering into
the states with $|k_x| \gtrsim 2k_F$, the equations (\ref{GEw})
decouple: for $k_x>0$ one has
 \begin{equation}
  \label{ginvPlus}
  \bpm
    i\omega - \eps_\sigma(\bfk) & -\Delta_{\bfQ\sigma} \\
    -\Delta_{-\bfQ\sigma} & i\omega - \eps_\sigma(\bfk-\bfQ)
  \epm
\hat{G} = \hat{I},
 \end{equation}
where
 \begin{equation}
  \label{hatG}
\hat{G}\equiv  \bpm
    g_\sigma(\bfk,\bfk,\omega) & g_\sigma(\bfk-\bfQ,\bfk,\omega) \\
    g_\sigma(\bfk,\bfk-\bfQ,\omega) & g_\sigma(\bfk-\bfQ,\bfk-\bfQ,\omega)
  \epm
 \end{equation}
and $\hat{I}$ is the $2\times 2$ identity matrix. The equation for
$k_x<0$ may be obtained via a substitution $\bfQ\to -\bfQ $.
Introducing the notations
 \begin{equation}
  \label{epspm}
  \eps^{\pm}_\sigma(\bfk',\bfk)=
    \frac{\eps_\sigma(\bfk')\pm \eps_\sigma(\bfk)}{2}
 \end{equation}
and using
$\Delta_{\bfQ\sigma}\Delta_{-\bfQ\sigma}=|\Delta_{\bfQ\sigma}|^2$
from (\ref{ginvPlus}) one has
\begin{widetext}
\begin{equation}
  \begin{split}
  \label{gkkminuQsCosine}
    g_\sigma(\bfk-\bfQ,\bfk,\omega)
    &=-\frac{\Delta_{\bfQ\sigma}}
    {[\omega+i\eps_\sigma(\bfk)][\omega+i\eps_\sigma(\bfk-\bfQ)]+
      |\Delta_{\bfQ\sigma}|^2}\\
    &=
    -\frac{\Delta_{\bfQ\sigma}}
    {[\omega+i\eps^+_\sigma(\bfk,\bfk-\bfQ)]^2
      +[\eps^-_\sigma(\bfk,\bfk-\bfQ)]^2
      +|\Delta_{\bfQ\sigma}|^2
    }.
  \end{split}
 \end{equation}
The consistency equation therefore is
 \begin{equation}
  \label{consistcosine}
  \Delta_{\bfQ\sigma'}=
    -T\sum_{\bfk\omega\sigma}
    \frac{(U_c-U_s\sigma\sigma') \Delta_{\bfQ\sigma}}
    { [\omega+i\eps^+_\sigma(\bfk,\bfk-\bfQ)]^2
      +[\eps^-_\sigma(\bfk,\bfk-\bfQ)]^2
      +|\Delta_{\bfQ\sigma}|^2}.
 \end{equation}
where $\omega $ takes the values $\pi T (2n+1),\ n\in Z$. Using the
identity
$$
  T\sum_\omega\frac{1}{(\omega-i\alpha)^2+a^2}=
  \frac{1}{4a}
  \left\{
    \th\frac{a-\alpha}{2T}
    +\th\frac{a+\alpha}{2T}
  \right\}
$$
this may be rewritten as
 \begin{equation}
  \label{consistcosineNoW}
  \Delta_{\bfQ\sigma'}=-\frac{1}{4}\sum_{\bfk\sigma}
  \frac{(U_c-U_s \sigma\sigma')\Delta_{\bfQ\sigma}}
  {\sqrt{|\Delta_{\bfQ\sigma}|^2+[\eps^-_\sigma(\bfk,\bfk-\bfQ)]^2}}
  \Bigl\{
    \th
    \frac{E_{\sigma,+}(\bfk )}{2T} -
    \th \frac{E_{\sigma,-}(\bfk )}{2T}
  \Bigr\},
 \end{equation}
where
 \begin{equation}
 \label{EnSpCos}
E_{\sigma,\pm}(\bfk )=\eps^+_\sigma(\bfk,\bfk-\bfQ)\pm
\sqrt{|\Delta_{\bfQ\sigma}|^2+[\eps^-_\sigma(\bfk,\bfk-\bfQ)]^2}
 \end{equation}
is the energy spectrum in the cosine phase. From (\ref{dispersion})
it follows that for the right-moving electrons
 \begin{align}
  \label{epsplusexpanded}
   \eps^+_\sigma(\bfk,\bfk-\bfQ)&=\frac{\hbar v_F q_x}{2} +
   2t_b\sin\frac{q_y b}{2}\sin\left( k_y b-\frac{q_y b}{2}\right)
   -2t'_b\cos q_y b\cos\left(2k_y b-q_y b\right) -\sigma H,\\
  \label{epsminusexpanded}
  \eps^-_\sigma(\bfk,\bfk-\bfQ)&=\hbar v_F(k_x-k_F-q_x/2) -
  2t_b\cos\frac{q_y b}{2}\cos\left( k_y b-\frac{q_y b}{2}\right)
  +2t'_b\sin q_y b\sin\left(2k_y b-q_y b\right).
 \end{align}

 It is, however, more convenient to perform the momentum summation
in Eq. (\ref{consistcosine}) first. Extending the limits of the
summation over $k_x$ to infinity, one has
 \begin{equation}
  \label{consistcosineSummedKX}
  \Delta_{\bfQ\sigma'}=\frac{\pi\nu_F |U_c|T}{2}\sum_{\omega\sigma}
   \left<
   \frac{(1+\nu \sigma\sigma')\Delta_{\bfQ\sigma}}
        {\sqrt{[\omega+i\eps^+_\sigma(\bfk,\bfk-\bfQ)]^2+
         |\Delta_{\bfQ\sigma}|^2 }}
   \right>_{k_y},
 \end{equation}
\end{widetext}
where the branch of the square root with positive real part is
implied, the angular brackets stand for the averaging over all
values of $k_y$, and we introduced the standard notation
 \begin{equation}
  \label{nu}
  \nu=-U_s/U_c
 \end{equation}
for the coupling ratio and
 \begin{equation}
  \label{nuF}
  \nu_F=\frac{L_x}{\pi\hbar v_F}
 \end{equation}
for the density of states on the Fermi level per one spin component.

Expansion to the third order in $\Delta_{\bfQ\sigma}$ yields
 \begin{equation}
  \label{consistcosine3rdOrder}
  \Delta_{\bfQ\sigma'}=\sum_\sigma\frac{1+\nu\sigma\sigma'}{2}
  \left( K^{(1)}_\sigma -|\Delta_{\bfQ\sigma}|^2  K^{(3)c}_\sigma
  \right) \Delta_{\bfQ\sigma}
 \end{equation}
(the superscript ``c'' stands for ``cosine'') with
 \begin{align}
  \label{K1cosine}
    K^{(1)}_\sigma &= \pi\nu_F |U_c|T\sum_\omega
    \left< \frac{\sign\omega}{\omega+i\eps^+_\sigma(\bfk,\bfk-\bfQ)} \right>_{k_y},\\
  \label{K3cosine}
    K^{(3)c}_\sigma &= \frac{\pi\nu_F |U_c|T}{2}\sum_\omega
    \left< \frac{\sign\omega}{(\omega+i\eps^+_\sigma(\bfk,\bfk-\bfQ))^3} \right>_{k_y}.
 \end{align}
The second-order transition line corresponds to
 \begin{equation}
  \label{transitionmatrix}
  \det
   \bpm
    \frac{1+\nu}{2}  K^{(1)}_+ - 1 &
    \frac{1-\nu}{2}  K^{(1)}_-  \\
    \frac{1-\nu}{2}  K^{(1)}_+  &
    \frac{1+\nu}{2}  K^{(1)}_- - 1 &
   \epm
   =0.
 \end{equation}
The left hand side of this equation is a function of $\nu ,T,H$ and
$\bfq$. The normal to CDW phase transition occurs at some optimal
value of $\bfq$ which corresponds to the minimum of the free energy
of the CDW state, that at the transition point is equivalent to the
maximum of $T_c$ or to the minimum of l.h.s. of Eq.
(\ref{transitionmatrix}).

Unrestricted summation over $k_x$ introduced divergence into the
summation over $\omega$ in $ K^{(1)}_\sigma$, which may be
renormalized by introducing the zero-field and zero-$t'_b$
transition temperature $T_{c0}$. To obtain a closed expression for
the latter, we note that for $H=0$ one has $ K^{(1)}_+= K^{(1)}_-$,
and the matrix in (\ref{transitionmatrix}) has eigenvalues $
K^{(1)}_\sigma-1$ and $\nu K^{(1)}_\sigma-1$. Therefore, at $H=0$
$K^{(1)}_\sigma(T=T_{c}(t'_b), H=0)=1$. From (\ref{K1cosine}) at
$t'_b=0$ after imposing a cutoff at $\omega \sim E_F$ one obtains
$K^{(1)}=\nu_F |U_c|\ln (2\gamma E_F/\pi T)$, and
 \begin{equation}
  \label{Tc0}
  T_{c0}=\frac{2\gamma E_F}{\pi} \exp \left(-\frac{1}{\nu_F |U_c|}\right) ,
 \end{equation}
in agreement with Ref. [\onlinecite{BZ99}] and with the usual BCS
expression. We may now write
\begin{widetext}
 \begin{equation}
  \label{K1cosineRenormalized}
  K^{(1)}_\sigma =
  1 + \nu_F |U_c|
  \left[
   \ln\frac{T_{c0}}{T}+\pi T\sum_{\omega}
    \left(
     \left< \frac{\sign\omega}{\omega+i\eps^+_\sigma(\bfk,\bfk-\bfQ )}\right>_{k_y} -
     \frac{\sign\omega}{\omega}
    \right)
  \right].
 \end{equation}
The kernels $K^{(1)}_\sigma$ and $K^{(3)c}_\sigma$ can be further
simplified in terms of the digamma function $\psi(x)=\frac{d}{dx}\ln
\Gamma(x)$:
 \begin{align}
  \label{K1psi}
  K^{(1)}_\sigma &=
  1 + \nu_F |U_c|
  \left[
   \ln\frac{T_{c0}}{T}+
     \psi\left(\frac{1}{2}\right)-
     \left< \Re\psi\left(\frac{1}{2}+\frac{i\eps^+_\sigma(\bfk,\bfk-\bfQ)}{2\pi
     T}\right) \right>_{k_y}
  \right],\\
  \label{K3cPsi}
  K^{(3)c}_\sigma &=
  -\frac{\nu_F |U_c|}{16\pi^2 T^2}
  \left< \Re \psi''\left(\frac{1}{2}+\frac{i\eps^+_\sigma(\bfk,\bfk-\bfQ)}{2\pi T}\right) \right>_{k_y}.
 \end{align}
To actually find the fields $\Delta_{\bfQ\sigma}$ just below the
transition point to the leading order, one has to make the
substitution $K^{(1)}_\sigma \to K^{(1)}_\sigma
-|\Delta_{\bfQ\sigma}|^2K^{(3)c}_\sigma$ in
(\ref{transitionmatrix}).  The ratio $\Delta_{\bfQ-} :
\Delta_{\bfQ+}$ is real and at the transition point is the column
ratio in (\ref{transitionmatrix}),
 \begin{equation}
 \label{columnratio}
 \Delta_{\bfQ-} : \Delta_{\bfQ+}
    =\left(\frac{1+\nu}{2}  K^{(1)}_+ - 1 \right) :
    \left(\frac{\nu-1}{2}  K^{(1)}_-  \right)
    =\left(\frac{\nu-1}{2}  K^{(1)}_+  \right) :
    \left(\frac{1+\nu}{2}  K^{(1)}_- -1  \right) .
 \end{equation}
Introducing the (real) ratio $\alpha \equiv |\Delta_{\bf Q-}|^2 :
|\Delta_{\bf Q+}|^2$ we get
 \begin{equation}
  \label{CosineDelta}
  |\Delta^{c}_{\bfQ\sigma}|^2=
  \frac{\alpha^{(1-\sigma )/2}
   \left(\nu K^{(1)}_+ K^{(1)}_-
    -\frac{\nu+1}{2}(K^{(1)}_+ +K^{(1)}_-)+1
   \right)}
  { K^{(3)c}_+ \left(\nu K^{(1)}_- -\frac{\nu+1}{2}\right)
  +\alpha K^{(3)c}_- \left(\nu K^{(1)}_+ -\frac{\nu+1}{2}\right)}.
 \end{equation}
\end{widetext}
The superscript ''c'' stands for the cosine phase. The CDW cosine
phase with $q_x\neq 0$ was analyzed at the transition line by means
of the susceptibility calculation and called the CDW$_x$ phase,
while the phase with $q_x=0$ was called CDW$_0$ \cite{ZBM}.

\subsection{The Double Cosine Phase}

Now we consider the solution with $\Delta_{\bfk\sigma} \neq 0$ for
$\bfk = \pm \bfQ_0 \pm \bfq$, where $\bfQ_0 = 2k_F\bfe_x+ (\pi/b)
\bfe_y$ and $q \ll k_F$.  Strictly speaking, there exist no
self-consistent solutions with only four harmonics present or all
others damped by a factor of $\Delta_{\bfQ\sigma}/E_F$.  However,
the modulations given above have equal second-order transition
temperatures, and immediately below the transition point other
harmonics are damped by a factor of $\Delta_{\bfQ\sigma}/T$ or
$\Delta_{\bfQ\sigma}/H$.

To obtain the leading-order expressions for the fields
$\Delta_{\bfQ\sigma}$ in the vicinity of the transition line, we
rewrite the equations (\ref{GEw}) in the ``matrix'' form
 \begin{equation}
 \label{EqGF}
  GG_0^{-1}-GF=I,
 \end{equation}
where
 \begin{equation}
 \label{GEmatrix}
 \begin{aligned}
  &G(\bfk',\bfk)=g_\sigma(\bfk',\bfk),\ \
  G_0(\bfk',\bfk)=\frac{\delta(\bfk',\bfk)}{i\omega-\eps_\sigma(\bfk)},\\
  &F(\bfk',\bfk)=\sum_Q \Delta_{\bfQ\sigma}
    \delta(\bfk',\bfk-\bfQ).
 \end{aligned}
 \end{equation}
The solution is
 \begin{equation}
 \label{Ex}
 G=G_0+G_0 F G_0 + G_0 F G_0 F G_0 + \ldots
 \end{equation}
Omitting the contribution from the virtual states with the momentum
$|k_x|\gtrsim 2k_F$, we obtain the consistency equation to the third
order in $\Delta_{\bfQ\sigma}$:
\begin{widetext}
 \begin{equation}
  \label{consistDcosine}
  \begin{split}
  \Delta_{\bfQ_0+\bfq,\sigma'}=&
  T\sum_{\bfk\omega\sigma}(U_c-U_s \sigma\sigma')\times
  \biggl[
   \frac{\Delta_{\bfQ_0+\bfq,\sigma}}
     {(i\omega-\eps_\sigma(\bfk))(i\omega-\eps_\sigma(\bfk-\bfQ_0-\bfq))}+\\
   &+\frac{\Delta_{\bfQ_0+\bfq,\sigma}|\Delta_{\bfQ_0+\bfq,\sigma}|^2}
     {(i\omega-\eps_\sigma(\bfk))(i\omega-\eps_\sigma(\bfk-\bfQ_0-\bfq))
      (i\omega-\eps_\sigma(\bfk))(i\omega-\eps_\sigma(\bfk-\bfQ_0-\bfq))}+\\
   &+\frac{\Delta_{\bfQ_0+\bfq,\sigma}|\Delta_{\bfQ_0-\bfq,\sigma}|^2}
     {(i\omega-\eps_\sigma(\bfk))(i\omega-\eps_\sigma(\bfk-\bfQ_0-\bfq))
      (i\omega-\eps_\sigma(\bfk-2\bfq))(i\omega-\eps_\sigma(\bfk-\bfQ_0-\bfq))}+\\
   &+\frac{\Delta_{\bfQ_0+\bfq,\sigma}|\Delta_{\bfQ_0-\bfq,\sigma}|^2}
     {(i\omega-\eps_\sigma(\bfk))(i\omega-\eps_\sigma(\bfk-\bfQ_0+\bfq))
      (i\omega-\eps_\sigma(\bfk))(i\omega-\eps_\sigma(\bfk-\bfQ_0-\bfq))}
  \biggr].
  \end{split}
 \end{equation}
The equation for the other two harmonics may be obtained via a
substitution $\bfq \to -\bfq$. Performing the integration over
$k_x$, we arrive at
 \begin{equation}
  \label{consistDcosine3rdOrder}
  \Delta_{\bfQ_0+\bfq,\sigma'}=
  \smash{\sum_\sigma\frac{1+\nu\sigma\sigma'}{2}}
  \Bigl( K^{(1)}_\sigma -|\Delta_{\bfQ_0+\bfq,\sigma}|^2 K^{(3)c}_\sigma
   -|\Delta_{\bfQ_0-\bfq,\sigma}|^2  K^{(3)d}_\sigma
  \Bigr) \Delta_{\bfQ_0+\bfq,\sigma}
 \end{equation}
(the superscript ``d'' stands for ``double cosine'') with $K^{(1)}$
and $K^{(3)c}$ given by (\ref{K1psi}) and (\ref{K3cPsi}), and
 \begin{equation}
  \begin{split}
  \label{K3d}
  K^{(3)d}_\sigma
  &= \pi\nu_F |U_c|T\sum_\omega
    \left< \frac{(\omega-i\sigma H)\sign\omega}
          {[(\omega+i\eps^+_\sigma(\bfk,\bfk-\bfQ))(\omega-i\eps^+_{-\sigma}(\bfk,\bfk-\bfQ))]^2}\right>_{k_y}=\\
  &=-\frac{\nu_F|U_c|}{4\pi\hbar v_F q_x T}
 \left< \left[
   \Im\psi'\left(\frac{1}{2}+\frac{i\eps^+_\sigma(\bfk,\bfk-\bfQ)}{2\pi T}\right)+
   \Im\psi'\left(\frac{1}{2}+\frac{i\eps^+_{-\sigma}(\bfk,\bfk-\bfQ)}{2\pi T}\right)
  \right]\right>_{k_y}.
  \end{split}
 \end{equation}
\end{widetext}
The function $K^{(3)d}_{\sigma}$ does not depend on $\sigma$ and
in what follows we omit this subscript.

At only a longitudinal shift of the CDW wave vector ($q_y=0$) the
$k_y$-dependence of $\eps^+_{-\sigma}(\bfk,\bfk-\bfQ)$ is symmetric,
and the functions $K^{(1)}_\sigma $, $K^{(3)c}_\sigma $ and
$K^{(3)d}$ (being dependent only on the shift wave vector $q_x$)
possess the symmetry: \begin{equation} \label{symK}
 \begin{split}
K^{(1)}_\sigma (q_x)&=K^{(1)}_{-\sigma} (-q_x), \\
K^{(3)c}_\sigma (q_x)&=K^{(3)c}_{-\sigma} (-q_x), \\
K^{(3)d} (-q_x)&=K^{(3)d} (q_x) .
 \end{split}
\end{equation} This symmetry follows directly from Eqs. (\ref{K1psi}),
(\ref{K3cPsi}) and (\ref{K3d}) and the properties of digamma
function: $\Re\psi^{(n)}\left( a+ib\right) = \Re\psi ^{(n)}\left(
a-ib\right) $, $\Im\psi^{(n)}\left( a+ib\right) = -\Im\psi
^{(n)}\left( a-ib\right) $. This symmetry is not a consequence of
the expansion in powers of $\Delta_{\sigma}$: the consistency
equation (\ref{consistcosineSummedKX}) also does not change under
the transformation \begin{equation} \label{sym} q_x\to -q_x, \sigma
\to -\sigma . \end{equation}

The ratio $|\Delta_{\bfQ_0+\bfq,\sigma}|^2
:|\Delta_{\bfQ_0+\bfq,-\sigma}|^2$ near the transition point is
still given by (\ref{columnratio}) accounting for the property
(\ref{symK}). Hence,
$|\Delta_{\bfQ_0-\bfq,\sigma}|^2:|\Delta_{\bfQ_0-\bfq,-\sigma}|^2 =
|\Delta_{\bfQ_0+\bfq,-\sigma}|^2 :|\Delta_{\bfQ_0+\bfq,\sigma}|^2
\equiv \alpha $. As in derivation of (\ref{CosineDelta}) one may
rewrite the system of equations (\ref{consistDcosine3rdOrder}) on
$\Delta $ as two equations (\ref{transitionmatrix}) with replacement
$K^{(1)}_{\sigma }(\bfQ_0 \pm \bfq_x) \rightarrow K^{(1)}_{\pm\sigma
}
   -|\Delta_{\bfQ_0 \pm \bfq,\sigma}|^2
    K^{(3)c}_{\pm\sigma }- |\Delta_{\bfQ_0\mp\bfq,\sigma }|^2  K^{(3)d}$.

Defining the ratio $\beta \equiv |\Delta_{\bfQ_0-\bfq,-}|^2 :
|\Delta_{\bfQ_0+\bfq,+}|^2$ we can extend the derivation of
(\ref{CosineDelta}) to get
\begin{widetext}
 \begin{equation}
  \label{DCosineDelta+}
  |\Delta_{\bfQ_0+\bfq,+}|^2=
  \frac{ \left(\nu K^{(1)}_+ K^{(1)}_-
    -\frac{\nu+1}{2}(K^{(1)}_+ +K^{(1)}_-)+1
   \right)}
  { \left(K^{(3)c}_+ + \alpha \beta K^{(3)d}\right)
      \left(\nu K^{(1)}_- -\frac{\nu+1}{2}\right)
  + \left(\alpha K^{(3)c}_- + \beta K^{(3)d}\right)
      \left(\nu K^{(1)}_+ -\frac{\nu+1}{2}\right)},
 \end{equation}
and from the $\bfQ_0-\bfq $ part of (\ref{consistDcosine3rdOrder})
 \begin{equation}
  \label{DCosineDelta-}
  |\Delta_{\bfQ_0-\bfq,-}|^2=
  \frac{\beta \left(\nu K^{(1)}_+ K^{(1)}_-
    -\frac{\nu+1}{2}(K^{(1)}_+ +K^{(1)}_-)+1
   \right)}
  { \left(\beta K^{(3)c}_+ +\alpha  K^{(3)d} \right)
      \left(\nu K^{(1)}_- -\frac{\nu+1}{2}\right)
  + \left( \alpha \beta K^{(3)c}_- + K^{(3)d} \right)
      \left(\nu K^{(1)}_+ -\frac{\nu+1}{2}\right)} .
 \end{equation}
Dividing (\ref{DCosineDelta+}) by (\ref{DCosineDelta-}) we obtain a
linear equation on $\beta $:
\begin{equation} \frac{\left(\beta K^{(3)c}_+
+\alpha  K^{(3)d}\right)
      \left(\nu K^{(1)}_- -\frac{\nu+1}{2}\right)
  + \left( \alpha \beta K^{(3)c}_- + K^{(3)d}\right)
      \left(\nu K^{(1)}_+ -\frac{\nu+1}{2}\right) }
{\left(K^{(3)c}_+ + \alpha \beta K^{(3)d}\right)
      \left(\nu K^{(1)}_- -\frac{\nu+1}{2}\right)
  + \left(\alpha K^{(3)c}_- + \beta K^{(3)d}\right)
      \left(\nu K^{(1)}_+ -\frac{\nu+1}{2}\right)}= 1,
\end{equation}
\end{widetext}
which gives $\beta = 1$. The values $\beta =0$ or $\beta =\infty$
in (\ref{DCosineDelta+}) and (\ref{DCosineDelta-}) correspond to
cosine phase. Hence, in the double cosine phase the symmetry
(\ref{sym}) is not broken, while the transition to cosine phase
breaks this symmetry.

\subsection{Free energy of cosine and double-cosine phases}

One can easily write down the expressions for the free energies of
cosine and double-cosine phases valid to the second order in the
energy gap $|\Delta_{\sigma}|$.

From (\ref{H}) we have for the free energy
 \begin{equation}
  \label{FcdwFn}
   F_{\rm CDW}-F_n=
   \frac{1}{2}\sum_{\bfQ\sigma} D_{-\bfQ\sigma}
   \Delta_{\bfQ\sigma}.
 \end{equation}
To the second order in $\Delta_{\bfQ\sigma}$ this rewrites as
 \begin{eqnarray}
  \label{FcdwFn2O}
   F_{\rm CDW}-F_n &=&-\frac{\pi \nu_F T}{4} \sum_{\bfQ\omega\sigma}
   \left<
    \frac{\Delta_{\bfQ\sigma} \Delta_{-\bfQ\sigma}\sign\omega}
         {\omega+i\eps^+_\sigma(\bfk,\bfk-\bfQ)}
   \right>_{k_y}\nonumber \\
   &=&-\frac{1}{4|U_c|}\sum_{\bfQ\sigma} K^{(1)}_\sigma
   |\Delta_{\bfQ\sigma}|^2.
 \end{eqnarray}
The phase with the most negative r.h.s of (\ref{FcdwFn}) wins;
positive values of the r.h.s. correspond to first-order transitions.
 From Eq. (\ref{FcdwFn2O}) using Eq. (\ref{symK}) we have
 \begin{equation}
  \label{FC}
   F_{c}-F_{n} =-\frac{ K^{(1)}_+ +\alpha K^{(1)}_-}{2|U_c|}
   |\Delta^{c}_{\bf Q_0+q_x,+}|^2
 \end{equation}
and
 \begin{equation}
  \label{F2C}
   F_{2c}-F_{n} =-\frac{K^{(1)}_+ +\alpha K^{(1)}_-}{|U_c|}
   |\Delta^{2c}_{\bf Q_0+q'_x,+}|^2 .
 \end{equation}
  The quantity $K^{(1)}_+ +\alpha K^{(1)}_- $ depends on $\bfq$ but is always
  positive near the metal-CDW transition line $T_c(H)$ where this transition
  is of the second kind.

Since CDW$_c$ and CDW$_{2c}$ phases have the same transition
temperature, the only way to determine which phase takes place is to
compare their free energies, that near the transition line $T_c(H)$
are given by the formulas (\ref{FC}) and (\ref{F2C}).
  The double cosine phase wins if the ratio
 \begin{equation}
  \label{RatioF}
  r_F \equiv \frac{F_{2c}-F_{n})}{(F_{c}-F_{n})} >1 ,
 \end{equation}
where the values of the functions $F_{2c}(T,H,\bfq )$ and
$F_{c}(T,H,\bfq )$ must be taken at the optimal value of the wave
vector $\bfq $, that should be found by minimization of these free
energy functions at each point of $T,H$ phase diagram and for each
of the two phases (CDW$_c$ and CDW$_{2c}$) separately.

 Below the transition temperature the optimal shift
vectors $\bfq$ could be different for the cosine and double cosine
phases (they minimize different free energy functions $F_{c}$ and
$F_{2c}$). However, at the transition temperature $T_c(H)$ the
CDW$_c$ and CDW$_{2c}$ phases have the same optimal value of $\bfq$
which is determined by the minimum of the left-hand side of Eq.
(\ref{transitionmatrix}). Hence, on the transition line $T_c(H)$ Eq.
(\ref{RatioF}) simplifies to
 \begin{equation}
  \label{RatioF1}
r_F=\frac{2|\Delta^{2c}_{\bf Q_0+q_x,+}|^2}{|\Delta^{c}_{\bf
Q_0+q_x,+}|^2}>1,
 \end{equation}
or after substitution of (\ref{CosineDelta}) and
(\ref{DCosineDelta+})
\begin{widetext}
 \begin{equation}
  \label{RatioF2}
r_F=\frac{2 \left[K^{(3)c}_+ \left(\nu K^{(1)}_-
-\frac{\nu+1}{2}\right)
  +\alpha K^{(3)c}_- \left(\nu K^{(1)}_+ -\frac{\nu+1}{2}\right) \right] }
{ \left(K^{(3)c}_+ + \alpha K^{(3)d}\right)
      \left(\nu K^{(1)}_- -\frac{\nu+1}{2}\right)
  + \left(\alpha K^{(3)c}_- + K^{(3)d}\right)
      \left(\nu K^{(1)}_+ -\frac{\nu+1}{2}\right)}>1.
\end{equation} \end{widetext} These formulas will be used later to determine
the phase diagram.

\section{The phase diagram}

In this section we only consider the longitudinal modulation
($q_y=0$) of the CDW wave vector. The CDW phase with $q_y\neq 0$
(CDW$_y$ phase) may appear for certain dispersion functions
$t_{\perp } (\bfk_{\perp })$. However, for the tight-binding model
(\ref{dispersion}) the CDW$_y$ phase is not expected to take place
near the transition temperature. Near the critical pressure (when
$t'_b\approx t'^*_b $ and the SDW-metal phase transition takes
place) in zero field the SDW$_2$ phase was predicted using the
susceptibility calculation from the normal state.\cite{HasFuk}
However, this phase was predicted only near $T=0$. Moreover, the
absolute instability line calculated in [\onlinecite{HasFuk}] does
not give the actual transition line at low temperature since this
transition takes place as the first-order phase transition.

To be more sure that we can disregard appearance of the CDW$_y$
phase in our consideration near the transition temperature we
performed the calculation of the optimal shift vector $\bfq $ at the
transition line $T_c(H)$ in a large range of parameters. We swept
the following three-dimensional range of parameters: $-0.3\leq \nu
\leq 0.9, 0\leq t'_b/t'^*_b<1$ and $0\leq \mu_B H\leq 2T_{c0}$ and
have not found that the shift of $Q_y$ leads to higher $T_c$
anywhere inside this range.

This does not contradict the prediction of the CDW$_y$ \cite{ZBM},
since according to this paper the CDW$_y$ phase appears at $\nu <0$
essentially due to the orbital effect of the magnetic field (see
Fig. 7d of Ref. [\onlinecite{ZBM}]). Below we will only consider the
range $0\leq \nu <1$ which is close to the experimental situation in
organic metals and disregard the appearance of CDW$_y$ phase.

\subsection{Transition line $T_c(H)$ and the normal-CDW$_0$-CDW$_x$ tricritical point}

The normal-to-CDW phase transition line is given by Eq.
(\ref{transitionmatrix}) irrespective of to which CDW phase this
transition occurs. The shift of the wave vector $q_x$ which enters
Eq. (\ref{transitionmatrix}) corresponds to the maximum value of
$T_c$ at given magnetic field $H$. The behavior of $T_c(H)$ and
$q_x(H)$ at the transition temperature has been analyzed in Ref.
[\onlinecite{ZBM}]. Diagonalization of the $M_3,M_4$ part of the
susceptibility matrix (6) in Ref. [\onlinecite{ZBM}] corresponds to
the diagonalization of the matrix in Eq. (\ref{transitionmatrix}).
On the transition line one of the eigenvalues of these matrices is
zero. In the case of perfect nesting equation
(\ref{transitionmatrix}) gives the same result for $T_c(H)$ as Eq.
(17) of Ref. [\onlinecite{ZBM}] (see fig. 7(a) of
[\onlinecite{ZBM}]).

For $q_x=0$ (CDW$_0$ phase) $K^{(1)}_{+}=K^{(1)}_{-}$ and Eq.
(\ref{transitionmatrix}) simplifies to $K^{(1)}_{+}=1$, that gives
for the transition temperature $T_c=T_c(H)$ the well-known equation
 \begin{eqnarray}
  \label{SimpleTc}
  \ln \left( \frac{T_{c}(H,t'_b)}{T_{c0}} \right) =
    \psi\left( \frac12 \right)  - \nonumber \\
 - \left< \Re\psi\left(\frac{1}{2}-\frac{i \eps^+_\sigma(\bfk,\bfk-\bfQ_0)}{2\pi T_{c}}\right)\right>_{k_y}
  ,
 \end{eqnarray}
where
 \begin{equation}
  \label{epsp0}
\eps^+_\sigma(\bfk,\bfk-\bfQ_0)=-2t'_b \cos (2k_y b) -\sigma H .
\end{equation}
 In
the Eq. (\ref{SimpleTc}) one can take either of the values $\sigma
=\pm 1$ since this equation does not depend on the sign of $\sigma
$. The transition line $T_c(H)$ from normal to CDW$_0$ phase does
not depend on the value of $\nu $.

The tricritical point (where the CDW$_0$ and CDW$_x$ phases have the
same transition temperature) is given by equation $\partial^2
D(\nu,T,H,q_x)/\partial q_x^2=0$, where $D(\nu,T,H,q_x)$ is the
left-hand side of Eq. (\ref{transitionmatrix}). At this point
$q_x=0, \Rightarrow \ K^{(1)}_{+}=K^{(1)}_{-}=1$ and $\partial ^n
K^{(1)}_{+}/\partial q_x ^n= (-1)^n \partial ^n K^{(1)}_{-}/\partial
q_x ^n$, and the equation for the tricritical point becomes
 \begin{equation}
\label{EqTri}
 \frac{\partial^2 K^{(1)}}{(\partial q_x)^2}=
 -\frac{2\nu }{1-\nu }\left(\frac{\partial K^{(1)}}{\partial q_x}\right)^2.
 \end{equation}
Substituting (\ref{K1psi}) and (\ref{K3cPsi}) this equation rewrites
as
 \begin{eqnarray}
 \label{EqTriPsi}
   -\left< \Re\psi ''\left(\frac{1}{2}-
   \frac{i \eps^+_{\uparrow }(\bfk,\bfk-\bfQ_0)}{2\pi
   T_{c}}\right)\right>_{k_y} =\nonumber \\
  \eta \left[
             \left< \Im\psi '\left(\frac{1}{2}-
   \frac{i \eps^+_{\uparrow }(\bfk,\bfk-\bfQ_0)}{2\pi T_{c}}\right)\right>_{k_y}
        \right]^2,
 \end{eqnarray}
where
 \begin{equation}
  \label{eta}
  \eta=2\nu\nu_F \vert U_c\vert /(1-\nu).
 \end{equation}
Together with Eq. (\ref{SimpleTc}) this equation allows to find the
tricritical point.

\subsection{$T_c(H)$ transition line at perfect nesting}

The case of perfect nesting ($t'_b=0$) is equivalent to the strictly
1D case in many mathematical aspects. However, the 3D nature of the
compound (its energy spectrum, e-e interaction and lattice
elasticity) preserves, and the mean-field approach still works. The
analysis in this subsection (at $t'_b=0$) is simple and performed
analytically. It helps to understand some qualitative features of
the CDW phase diagram in magnetic field.

At perfect nesting and purely longitudinal modulation of CDW
($q_y=0$) the expressions for $K^{(1)}_\sigma$, $K^{(3)d}$ and
$K^{(3)c}_\sigma$ simplify to
\begin{widetext}
 \begin{align}
  \label{K1psi1D}
  K^{(1)}_{\sigma 0}&=
  1 + \nu_F |U_c|
  \left[
   \ln\frac{T_{c0}}{T}+
     \psi\left(\frac{1}{2}\right)-
     \Re\psi\left(\frac{1}{2}+\frac{ih_\sigma}{2\pi T}\right)
  \right], \\
  \label{K3d1D}
  K^{(3)d}_0 &=-\frac{\nu_F|U_c|}{4\pi\hbar v_F q_x T} \left[
   \Im\psi'\left(\frac{1}{2}+\frac{ih_\sigma}{2\pi T}\right)+
   \Im\psi'\left(\frac{1}{2}+\frac{ih_{-\sigma}}{2\pi T}\right)
  \right],
 \end{align}
\end{widetext}
 and
  \begin{equation}
  \label{K3cPsi1D}
  K^{(3)c}_{\sigma 0}=
  -\frac{\nu_F |U_c|}{16\pi^2 T^2}
  \Re \psi''\left(\frac{1}{2}+\frac{ih_\sigma}{2\pi T}\right) ,
 \end{equation}
 where
 \begin{equation}
  \label{hsigma}
  h_\sigma = \frac{\hbar v_F q_x}{2}-\sigma H.
 \end{equation}

At $q_x=0$, Eq. (\ref{K3d1D}) even more simplifies:
 \begin{equation}
  \label{K3ds}
  K^{(3)d}_{00} = -\frac{\nu_F |U_c|}{8\pi^2 T^2}
  \Re \psi''\left(\frac{1}{2}-\frac{i H}{2\pi T}\right) \\
  =2K^{(3)c}_{\sigma\, 00}.
 \end{equation}

One can now derive a simple formula for the transition line $T_c(H)$
in the high field limit $H\gg T_c(H)$. As $H\to \infty $, $ h_\sigma
$ in (\ref{hsigma}) goes to zero for one spin component and to
$-2\sigma H $ for the other. Using the limit expansion of the
digamma function, $\Re\psi(1/2+ix)=\ln x +O(1/x),\ x\to\infty$, from
(\ref{K1psi1D}) we have at $H/\pi T_{c}(H)\gg 1$
 \begin{equation}
  \begin{split}
  \label{K1Lim}
   K^{(1)}_{+} &\approx 1 + \nu_F |U_c| \ln (T_{c0}/T) \\
   K^{(1)}_{-}&\approx
    1+\nu_F |U_c|\ln\left( \pi T_{c0}/4\gamma H\right) .
  \end{split}
 \end{equation}
Equation (\ref{transitionmatrix}) rewrites as
$$
K^1_+=\frac{(1+\nu) K^{(1)}_-/2 -1}{\nu K^{(1)}_- -(1+\nu)/2},
$$
that in the limit $H/\pi T_{c}(H)\gg 1$ becomes
 \begin{equation}
  \label{LogTcH}
   \ln (T_{c0}/T)=\frac{\ln (4\gamma H/\pi T_{c0}) }
   {\eta \ln(4\gamma H/\pi T_{c0})+1}.
 \end{equation}
At $\eta\ln (4\gamma H/\pi T_{c0}) \gg 1 $ this simplifies to
 \begin{equation}
  \label{TcH}
   T_c(H\to \infty)= T_{c0} \exp\left(-1/\eta \right).
 \end{equation}
In the intermediate interval $\pi T_{c}(H) \ll 4\gamma H \ll \pi
T_{c0} \exp (1/\eta )$ we get
 \begin{equation}
  \label{TcH1}
   T_c(H)\approx \pi T_{c0}^2/(4\gamma H).
 \end{equation}
Hence, at perfect nesting in the limit $H\to \infty$ the transition
temperature tends to a finite value which is natural since it
corresponds to the CDW instability for only one spin component.

However, the behavior of $T_c(H)$ given by formulas (\ref{LogTcH}),
(\ref{TcH}) and (\ref{TcH1}) strongly changes at finite
''antinesting'' term $t'_b$ (see below).

\subsection{Metal-CDW transition lines and tricritical points at finite $t'_b$}

\begin{figure}
\includegraphics{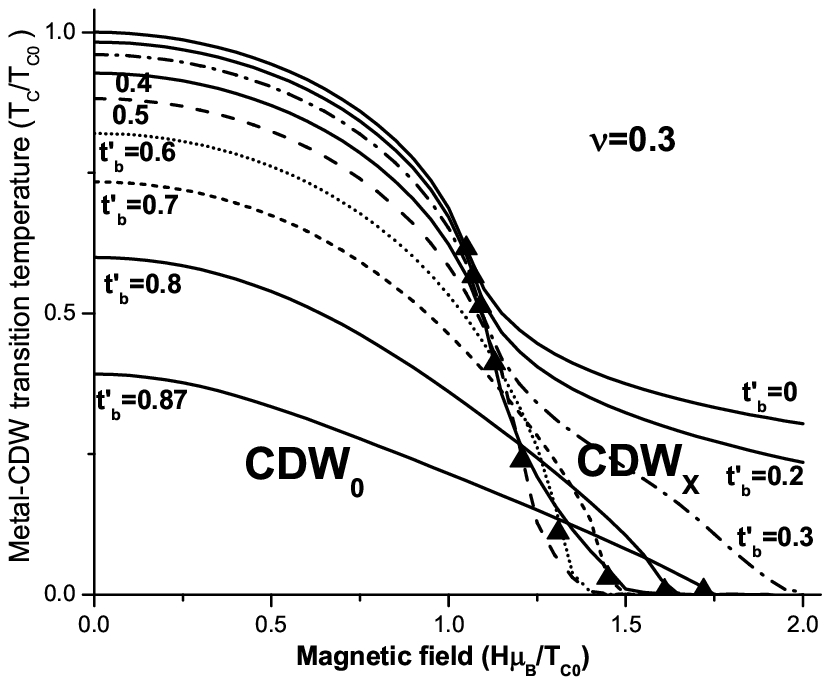}
\includegraphics{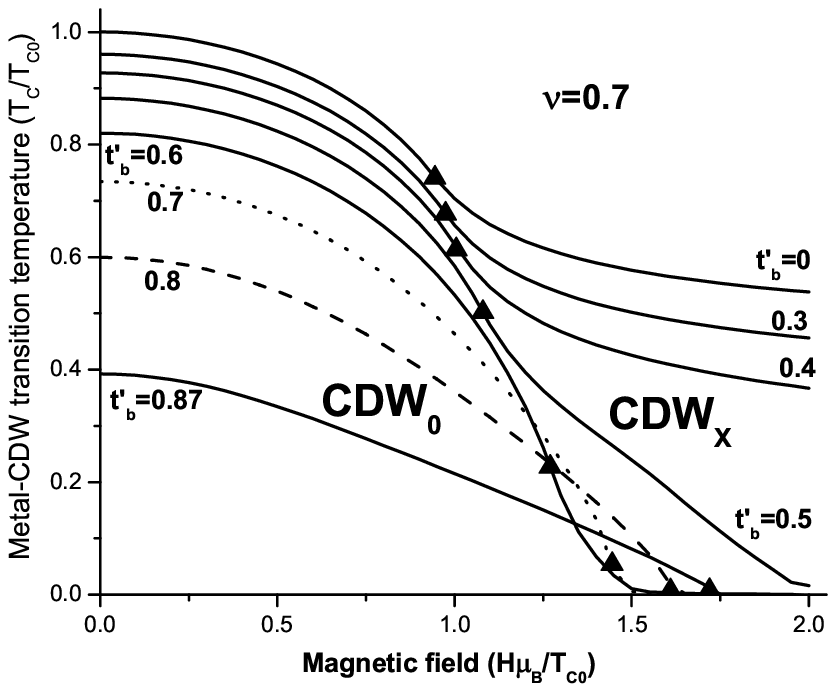}
\includegraphics{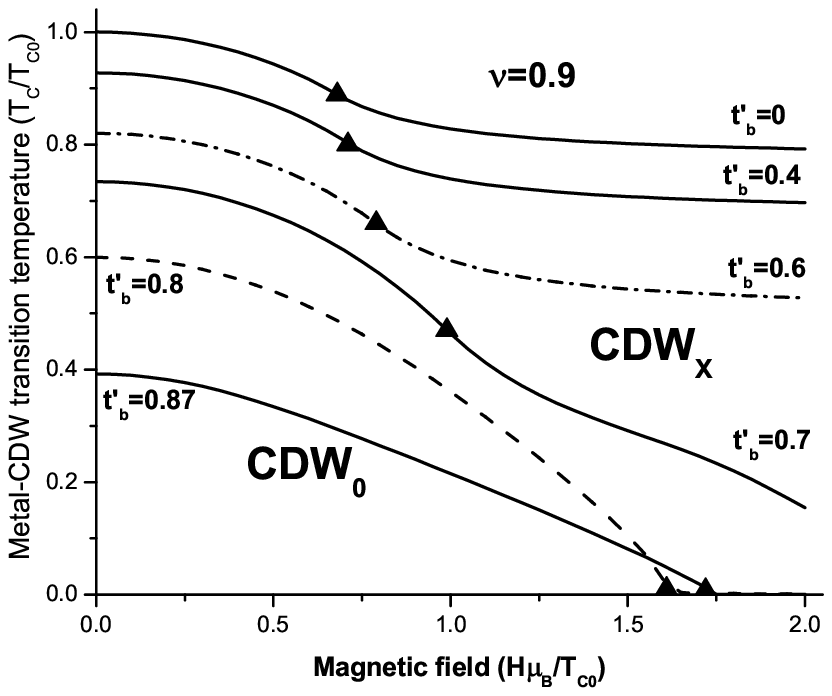}
 \caption{\label{TcHtb} Transition lines $T_c(H)$ at
 different values of $t'_b$
 at three different values of the coupling constant ratio $\nu$.
The values of $t'_b$ is given on the figures in units of $T_{c0}$;
for the tight-binding dispersion (\ref{dispersion}) the critical
value of $t'_b$ at which CDW disappears without magnetic field is
$t'^*_b=\Delta_0/2\approx 0.88T_{c0}$. The triangles show the
tricritical points Normal-CDW$_0$-CDW$_x$. }
\end{figure}

Using the Eq. (\ref{transitionmatrix}) and Eq. (\ref{EqTriPsi}) we
calculated the transition lines $T_c(H)$ and the tricritical points
at different values of $t'_b$ and $\nu $. The results are shown in
Fig. \ref{TcHtb}. From this figure we see that the transition line
$T_c(H)$ depends strongly on the both parameters $t'_b$ and $\nu $,
and one cannot determine the value of $t'_b$ from the $T_c(H)$
transition line if one does not know the value of $\nu $. Hence, to
determine experimentally what values take $t'_b$ and $\nu $ in a
particular compound, one has to perform two independent tests. For
example, one can study the pressure dependence of the $T_c(H)$ lines
at two different tilt angles of magnetic field: the first
measurement when the magnetic field is parallel to the conducting
$x$-$y$ plane and the formulas derived above in neglect of orbital
effects are valid, the second measurement in the very strong
magnetic field perpendicular to the conducting $x$-$y$ plane, when
the orbital effects are so strong that the one-dimensionization of
the electron dispersion takes place and the formulas for perfect
nesting (see sec. 3.2) become approximately valid. One can also
determine the value of $t'_b$ independently from the transport
measurement and then compare it to the critical value of
$t'^*_b=\Delta_0/2\approx 0.88T_{c0}$ at which the CDW state is
damped without magnetic field.

There is one interesting common feature on all diagrams in Figs.
\ref{TcHtb}. At each value of $\nu $ there is a critical value of
$t'_b<t'^*_b$ above which the CDW$_x$ phase disappears, i.e. as
magnetic field increases the transition from CDW$_0$ to metal state
instead of the CDW$_x$ phase takes place up to the lowest
temperature. This critical value of $t'_b$ may shift due to the
''one-dimensionization'' effect of magnetic field.

We have also checked if the region of cosine phase increases
considerably after finite ''antinesting'' term is taken into
account. The results of this study at the transition line are given
in Fig. \ref{Hc12Nu}. These results indicate that the region of
cosine phase does not increase considerably at finite $t'_b$ for the
tight-binding dispersion.

\subsection{Cosine and double-cosine phases at perfect nesting}

To determine which of the two phases (CDW$_c$ or CDW$_{2c}$) wins we
have to compare their free energies given in section 2.5.

On the transition line this reduces to the evaluation of the ratio
(\ref{RatioF2}).
 At the tricritical point (when $q_x=0$)  this
 ratio can be easily evaluated using Eq. (\ref{K3ds}):
 \begin{equation}
  \label{RatioFTri}
  r_{Ftriple}=2/3<1.
 \end{equation}
Hence, near the tricritical point the cosine phase wins. In the
limit of high field ($H\gg T_c$) one has $\hbar v_F q_x \to 2H$ and
$h_\sigma \to (1-\sigma )H$. In this limit $K^{(3)c}_+ \approx
-(\nu_F |U_c|/16\pi^2 T^2)
  \Re \psi''\left( 1/2\right)\gg K^{(3)c}_-, K^{(3)d}$, and
from (\ref{RatioF2}) we get
 \begin{equation}
  \label{RatioHigh}
 r_F\to 2 \ {\rm at}\  H/T_c \to \infty .
 \end{equation}
Hence, at high magnetic field the double cosine phase wins. These
two simple estimates suggest that the cosine and double-cosine
phases both appear on the phase diagram.

The boundary between CDW$_c$ and CDW$_{2c}$ phases on the transition line $T_c(H)$
is given by the equation $r_F(T,H,\nu)=1$, which rewrites as
 \begin{equation}
  \label{c2cBoundary}
 \begin{split}
 \left( K^{(3)c}_+ - \alpha K^{(3)d}\right)
      \left(\nu K^{(1)}_- -\frac{\nu+1}{2}\right) =\\
 = \left( K^{(3)d} - \alpha K^{(3)c}_- \right)
      \left(\nu K^{(1)}_+ -\frac{\nu+1}{2}\right) .
\end{split} \end{equation}
This equation is valid also for nonzero $t'_b$ and together with Eq.
(\ref{transitionmatrix}) allows to determine the second tricritical
point $(H_{c2}(\nu ), T_c(H_{c2}))$ where the normal, CDW$_c$ and
CDW$_{2c}$ phases meet.

\begin{figure}
\includegraphics{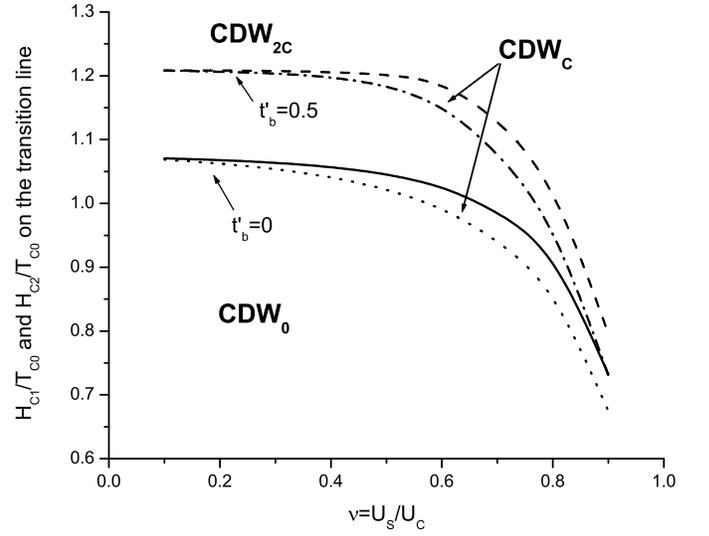}
\caption{\label{Hc12Nu} The magnetic field values $H_{c1}(\nu )$ and
$H_{c2}(\nu )$ at two tricritical points Normal-CDW$_0$-CDW$_c$ and
Normal-CDW$_c$-CDW$_{2c}$ correspondingly as function of coupling
constant ratio $\nu =U_s/U_c$ for two different values of $t'_b$.
The solid and dot lines denote $H_{c2}(\nu )$ and $H_{c1}(\nu )$ at
perfect nesting ($t'_b=0$) while the dash and dash-dot lines denote
$H_{c2}(\nu )$ and $H_{c1}(\nu )$ at finite value of the
''antinesting'' term ($t'_b=0.5T_{c0}$). The cosine phase exists in
the areas between $H_{c1}(\nu )$ and $H_{c2}(\nu )$ lines at the
same value of $t'_b$. }
\end{figure}

In Fig. \ref{Hc12Nu} we plot two triple-point values $H_{c1}(\nu )$
and $H_{c2}(\nu )$ of magnetic field which are given by equations
(\ref{EqTriPsi}) and (\ref{c2cBoundary}) correspondingly (together
with Eq. (\ref{transitionmatrix})) at two different values of
$t'_b=0$ and $t'_b=0.5T_{c0}$. $H_{c1}(\nu )$ and $H_{c2}(\nu )$
determine the tricritical points Normal-CDW$_0$-CDW$_c$ and
Normal-CDW$_c$-CDW$_{2c}$. From this figure we see that the range of
the cosine phase is very narrow (the difference $H_{c2}-H_{c1}$ does
not exceed 5\% of $H_{c1}$) and strongly depends on $\nu$.

At perfect nesting two tricritical points $H_{c1}$ and $H_{c2}$
coincide at $\nu =0$. At this "double-triple" point the denominator
in formulas (\ref{CosineDelta}) and (\ref{DCosineDelta+}) is zero.
This means, that one should expand up to the higher orders in
$|\Delta^{2}|$ than fourth and that at this point the critical
fluctuations are very strong.

From Fig. \ref{Hc12Nu} we see that (i) on the phase diagram of CDW
in magnetic field there are at least 3 different CDW states:
CDW$_0$, CDW$_c$ and CDW$_{2c}$, (ii) at the transition line
$T_c(H)$ the region of the CDW$_c$ phase is quite narrow for various
values of $t'_b$ and is usually sandwiched between the CDW$_0$ and
CDW$_{2c}$ phases as the magnetic field increases.

To analyze how this picture evolves below the transition line
$T_c(H)$ we performed a numerical calculation of the free energies
of CDW$_{c}$ and CDW$_{2c}$ phases using formulas
(\ref{FC}),(\ref{F2C}),(\ref{DCosineDelta+}),(\ref{CosineDelta}).
These formulas are valid if $\Delta \ll \pi T,H$, that covers a
narrow region below $T_c(H)$.

\begin{figure}
\includegraphics{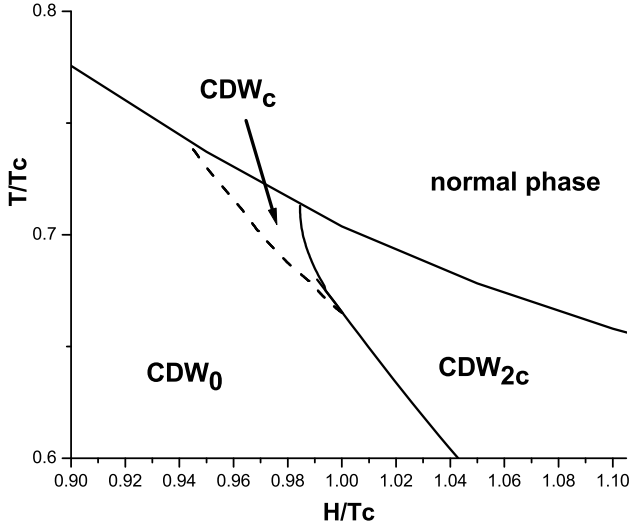}
\caption{\label{PhDTri} Phase diagram in $H-T$ coordinates at $\nu
=0.7$ at perfect nesting near the tricritical point. The CDW$_c$
region disappears rapidly as temperature decreases below $T_c$.}
\end{figure}

The computation performed in the case of perfect nesting shows that
the region of the CDW$_c$ phase becomes even more narrow as the
temperature decreases and disappears at $T\approx 0.95 T_c(H)$ (see
Fig. \ref{PhDTri}). Hence, the cosine phase at perfect nesting
exists only in a very small region of the phase diagram where it is
strongly smeared by the critical fluctuations.

The similar phase diagram appears in surface superconductors in
parallel magnetic field.\cite{Dimitrova} The cosine and
double-cosine CDW phases correspond to the helical and cosine
(stripe) superconducting phases respectively, and the total CDW$_x$
phase corresponds to the non-uniform superconducting LOFF
state.\cite{FFLOFF,FFLOLO} However, our situation differs from that
 in Ref. [\onlinecite{Dimitrova}], where the Rashba term in
the electron dispersion relation plays an important role in
obtaining the phase diagram similar to that on Fig. \ref{PhDTri}.
The number of harmonics and coupling constant in our case is twice
larger than that in Ref. [\onlinecite{Dimitrova}]. However, the
analogy between CDW and surface superconductors in magnetic field is
rather deep (see the discussion section).

\begin{figure}
\includegraphics{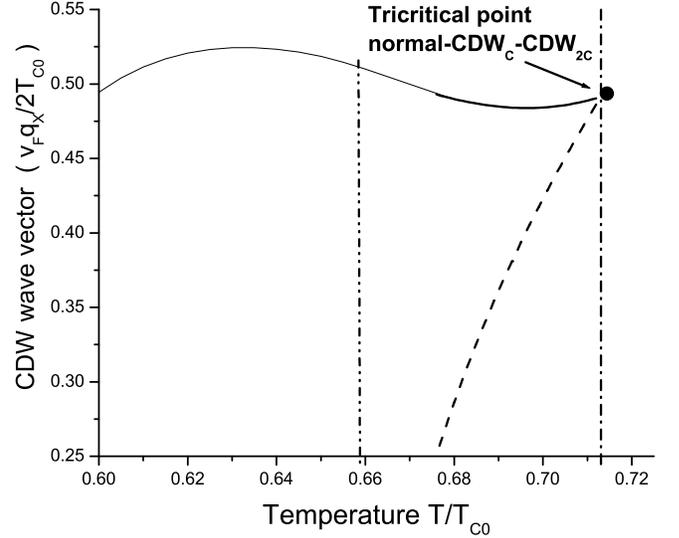}
\caption{\label{QxH} The shifts $q_x$ of wave vectors of the phases
CDW$_{2c}$ (solid line) and CDW$_c$ (dash line) along the
CDW$_c$-CDW$_{2c}$ transition line (at $0.713>T/T_{c0}>0.66$) or
CDW$_{0}$-CDW$_{2c}$ transition line (at $T/T_{c0}<0.66$) as
function of temperature. These shifts coincide only at the
normal-CDW$_c$-CDW$_{2c}$ tricritical point at $T/T_{c0}=0.713$. The
difference between the solid and dash lines gives the jump of the
CDW wave vector on the CDW$_c$-CDW$_{2c}$ transition line that makes
this transition of the first kind and leads to hysteresis.The
dash-dot line stands for the normal-CDW transition temperature, and
the dash-double-dot line shows the temperature of
CDW$_{0}$-CDW$_c$-CDW$_{2c}$ tricritical point.}
\end{figure}

On the transition line CDW$_c$-CDW$_{2c}$ the energy gaps and the
optimal shifts $q_x$ of the CDW wave vectors of these two phases
become substantially different as the temperature decreases below
$T_c$ (see Figs. \ref{QxH}). This means that the transition from
CDW$_c$ to CDW$_{2c}$ is of the first order. The transition line
between CDW$_0$ and CDW$_{2c}$ is also of the first order, and the
CDW$_{2c}$ nesting vector differs from $Q_0=2k_F$ already on the
transition line (see Fig. \ref{QxH}). This fact explains the huge
hysteresis in magnetization and magnetoresistance observed in
$\alpha$-(BEDT-TTF)$_2$KHg(SCN)$_4$ at the kink
transition.\cite{Qualls,Andres,OrbQuant} The first-order phase
transition driven by magnetic field is usually accompanied by
hysteresis. The jump of the nesting vector on the CDW$_0$-CDW$_{2c}$
transition line itself leads to huge hysteresis. The CDW and its
wave vector are pinned by the impurities and crystal
imperfections.\cite{Gruner} As magnetic field increases through the
kink transition point $H=H_c(T)$, the jump of the CDW wave vector
forces the CDW condensate to move. Due to the pinning this motion is
hysteretic since at each time moment the CDW condensate finds some
local minimum of the impurity potential. The larger jump of the CDW
wave vector at the kink transition the greater the hysteresis is.
This jump decreases with increasing temperature and comes to zero on
the transition line $T_c(H)$. The hysteresis shows a similar
temperature dependence. The hysteresis reduces with heating also due
to the thermal activation processes which reduce the pinning.

\section{Discussion}

 The electron charge and spin densities and the lattice distortion
 differ considerably in the cosine CDW$_c$ and double-cosine
 CDW$_{2c}$ phases that can help to
distinguish these two states experimentally. In the double-cosine
phase the charge modulation is the sum of two cosine distortions
that leads to the beats of the charge density wave:
\begin{eqnarray}
\label{rhoch2c} \rho_C^{2c} (x)=\nu_F (\Delta_{+}+\Delta_{-})\{ \cos
[(Q_0+q_x)x+\phi_1]
\nonumber \\
+\cos [(Q_0-q_x)x+\phi_2]\} \nonumber \\
=2\nu_F (\Delta_{+}+\Delta_{-})\cos [Q_0 x+(\phi_1+\phi_2)/2]\nonumber \\
\times \cos [q_x x+(\phi_1-\phi_2)/2].
\end{eqnarray}
The phase shifts $\phi_1$ and $\phi_2$ may depend on coordinate
because of the pinning of CDW by impurities. Usually,
$\Delta_{+}(Q)$ and $\Delta_{-}(Q)$ in CDW$_{2c}$ state differ
substantially, and the charge density wave in magnetic field is
accompanied by a spin density wave. The spin-density modulation in
the double-cosine phase is given by
\begin{eqnarray}
\label{rhoS2c} \rho_S^{2c} (x)=\nu_F (\Delta_{+}-\Delta_{-}) \{\cos
[(Q_0+q_x)x+\phi_1] \nonumber \\
- \cos [(Q_0-q_x)x+\phi_2]\}\nonumber \\
=-2\nu_F (\Delta_{+}-\Delta_{-})\sin [Q_0 x+(\phi_1+\phi_2)/2]
\nonumber
\\
\times \sin [q_x x+(\phi_1-\phi_2)/2].
\end{eqnarray}
In the cosine phase both charge and spin densities have one-cosine
modulations,
\begin{eqnarray}
\label{rhocos} \rho_C^{c} (x)=\nu_F (\Delta_{+}+\Delta_{-}) \cos
[(Q_0+q_x)x+\phi_1] \nonumber \\
\rho_S^{c} (x)=\nu_F (\Delta_{+}-\Delta_{-}) \cos
[(Q_0+q_x)x+\phi_1].
\end{eqnarray}
Thus, the charge modulations in the cosine and double-cosine phases
can be experimentally distinguished by X-ray or Raman scattering,
and the spin modulation can be detected by muon or neutron
scattering experiments.

The energy spectrum in the cosine and double-cosine phases also
differ strongly. In the cosine phase the energy spectrum
(\ref{EnSpCos}) is asymmetric with respect to the spin components
since the energy gaps $\Delta_{\sigma}$ differ for two spin
components $\sigma$. This means that under external electric field
the spin current is produced in addition to the charge current since
the charge is transferred by electrons with predominantly one spin
component. The degree of current polarization depends on the shift
of CDW wave vector, and, hence, can be controlled by the external
magnetic field. This property of the cosine phase may find
applications in spintronics. The double-cosine phase is symmetric in
spin components, and its energy spectrum has at least two gaps for
each spin component. The symmetry (\ref{sym}) preserves in the
double-cosine phase while in the cosine phase it is spontaneously
broken. Thus, the one-cosine and double-cosine CDW states differ
substantially in their thermodynamic, transport and optical
properties.

At low temperature $T\ll T_c$ the expansion in (\ref{Ex}) is not
applicable and the additional harmonics ${\bf Q} +(2n+1){\bf q_x}$
with integer $n$ appear in the double-cosine solution of the
consistency equation. These harmonics make the charge modulation in
the CDW$_{2c}$ phase at low temperature essentially nonsinusoidal in
space and, possibly, containing soliton walls. At low temperature
the quasi-particles differs substantially from the pair excitations
with activation energy $2\Delta $ suggested by the simple mean-field
description. Even at finite inter-chain electron dispersion the
variational methods using the soliton solutions of the consistency
equations show \cite{BrazKirovaReview} that the lowest-energy
excitations do not resemble the one-electron-type quasi-particles
but involve many electrons and are accompanied by a ''soliton''
kinks in the coordinate dependence of the order-parameter
$\Delta_{\sigma }(x)$. These excitations in the incommensurate CDW
have spin $1/2$, zero charge and the energy $2\Delta /\pi$ for a 1D
chain without magnetic field. In external magnetic field their
activation may become energetically favorable at $H>2\Delta_0 /\pi$
\cite{BDK} that corresponds to the phase-transition from CDW$_0$ to
the CDW$_x$ phase. Unfortunately, there is no a complete solution of
this problem at nonzero temperature at present time. The calculation
at the electron density close to half-filling \cite{Machida} does
not describe properly the incommensurate case far from the
half-filling. In particular, it considers only the one-cosine
modulation of the charge density. As we have seen, it is usually
energetically less favorable than the double-cosine phase in high
magnetic field. Therefore, we do not present an exact phase diagram
in the whole temperature and magnetic field range.

\begin{figure}
\includegraphics{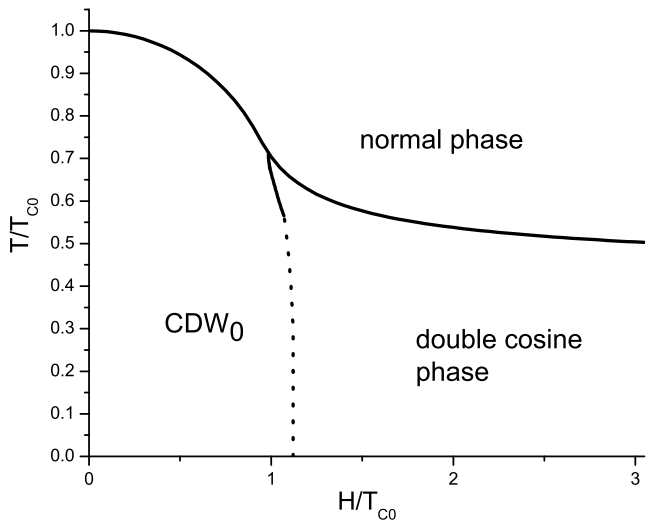}
\caption{\label{PhDTot} Phase diagram in $H-T$ coordinates at
perfect nesting in the whole essential region of CDW at $\nu =0.7$.
The solid lines have been calculated from Eqs.
(\ref{transitionmatrix}), (\ref{FC}),(\ref{F2C}),(\ref{CosineDelta})
and (\ref{DCosineDelta+}). The CDW$_c$ region is so narrow that it
can hardly be distinguished on the total phase diagram. The dot line
is a smooth link between the CDW$_0$-CDW$_{2c}$ boundary near $T_c$
and its value $H_c=2\Delta_0/\pi \approx 1.12T_{c0}$ at $T=0$.}
\end{figure}

Qualitatively, for the tight-binding dispersion one may expect the
following picture. The cosine phase does not appear on the phase
diagram besides the small region near the tricritical point as shown
on Fig. \ref{PhDTri}. One can easily show that in the case of
perfect nesting at $T\to 0$ the cosine phase is always unstable
toward the formation of the double-cosine modulation. However, an
observation of the cosine phase would be very interesting because of
its spin asymmetry in thermodynamic and transport properties. For
this reason some further calculation of the phase diagram for
various electron dispersion relations would be interesting.
 The transition from CDW$_{0}$ to CDW$_{2c}$ phase at low temperature
comes about as the appearance of soliton kinks (or soliton walls).
\cite{BDK} The soliton phase described in Ref. [\onlinecite{BDK}]
has two forbidden bands (energy gaps) which makes it similar to the
double-cosine phase. With
 increase of magnetic field the density of kinks increases, and the order
parameter comes continuously to the sinusoidal modulation close to
that in the double-cosine phase. At perfect nesting the
CDW$_0$-CDW$_{2c}$ boundary may be obtained qualitatively at
arbitrary temperature as a smooth link between Fig. \ref{PhDTri} and
the point $H_c=2\Delta_0/\pi\approx 1.1T_{c0}$ (see Fig.
\ref{PhDTot}).

In the case of non-uniform LOFF state the susceptibility calculation
in metal phase only gives the length of the optimal wave vector
$|\boldsymbol{q}_{opt}|$ of the order-parameter modulation. To
determine the most energetically favorable combination of
$\boldsymbol{q}:\ |\boldsymbol{q}|=|\boldsymbol{q}_{opt}|$ one has
to consider the third-order terms in $|\Delta |$ in the consistency
equation. To determine which phase, CDW$_c$ or CDW$_{2c}$, wins we
performed the similar procedure. Our procedure takes into account
some peculiarities of the CDW state such as the influence of the
''antinesting'' harmonic in the electron dispersion.

We have already mentioned the resemblance of the phase diagram
obtained for CDW (Fig. \ref{PhDTri}) with that of layered
superconductors with Rashba term.\cite{Dimitrova} The
CDW$_0$-CDW$_{2c}$ transition at low temperature also has many
common features with the transition from uniform superconducting
state to LOFF state in layered superconductors.\cite{BurkRainer} In
that case at the first critical field $H_{c1}$ the formation of a
single soliton kink also becomes energetically favorable, and the
functional shape of this soliton is well approximated by $\Delta
(x)=\Delta_0 \tanh (x/x_0)$ both in CDW$_{2c}$ and LOFF states at
low temperature. However, there are some important differences
between these two systems. In the case of superconductivity the
phase transition to the LOFF state was predicted to be of the second
kind \cite{BurkRainer}, and the concentration of soliton walls
increases gradually from zero with the increase of the magnetic
field. In Ref. [\onlinecite{BDK}] the first-order phase transition
was predicted between the uniform CDW and soliton phases. Our
calculation also suggests the first-order transition between CDW$_0$
and CDW$_{2c}$ phases. The first-order phase transition from metal
to the LOFF phase was predicted by both the mean-field
\cite{BurkRainer} and renormalization-group studies.\cite{KunYang}
The renormalization group analysis suggests \cite{KunYang} that the
enhanced fluctuations of the LOFF state, associated with additional
broken symmetries, are important to make this transition of the
first order. We do not consider the effect of fluctuations.
According to the criterion in Refs.
[\onlinecite{HorGutWeger,McKenzieFluct}] the effects of fluctuations
must not destroy the mean-field solution in organic metals where the
interlayer transfer integral $t_b$ is, usually, much larger than
$T_{c0}$. The transition from the normal state to the CDW$_x$ phase
observed on experiments
\cite{KartsLaukin,Biskup,Christ,Qualls,Andres,HarrCDW,OrbQuant,UncCDW1,UncCDW2,Graf,Graf1}
seems also to be of the second order. Further analysis of the
CDW$_0$-CDW$_x$ phase diagram in the case of imperfect nesting at
other dispersion relation can give more accurate description and
even new qualitative features.

Our results are aimed to help to analyze the rather complicated
phase diagram of CDW state in high magnetic field observed in a
number of organic metals, such as
$\alpha$-(BEDT-TTF)$_2$KHg(SCN)$_4$
\cite{KartsLaukin,Biskup,Christ,Qualls,Andres,HarrCDW,OrbQuant,UncCDW1,UncCDW2}
(see also a discussion in [\onlinecite{McKenzieCondMat}]),
(Per)$_2$M(mnt)$_2$ (M being Au,Pt,Cu) \cite{Graf,Graf1,Per1} etc,
where the transition from CDW$_0$ to high-field CDW$_x$ phase is
within the experimentally achievable magnetic fields. The above
investigation may also be applied to analyze the phase diagram in
nonorganic compounds with CDW ground state \cite{Monceau} (the
attainable magnetic field in pulsed magnets is already comparable to
the transition temperatures in some of these compounds). An accurate
quantitative description and comparison with the experimental
observations in Refs.
[\onlinecite{KartsLaukin,Biskup,Christ,Qualls,Andres,HarrCDW,OrbQuant,UncCDW1,UncCDW2,Graf,Graf1}]
may require the substitution of a more realistic electron dispersion
relation for each compound in the formulas derived above. With the
two-harmonic tight-binding dispersion our study already explains
several qualitative features of the CDW state in high magnetic
field. First, we show that the transition from CDW$_0$ to CDW$_x$
state is of the first kind and is accompanied by the substantial
jumps of the energy gap in electron spectrum and of the nesting
vector that results in strong hysteresis. This fact being observed
in many experiments on $\alpha$-(BEDT-TTF)$_2$KHg(SCN)$_4$ has not
received a theoretical substantiation before. Second, we propose and
prove the appearance of double-cosine CDW above the kink field which
is consistent with the low temperature soliton solution. Third, we
describe some properties of the cosine and double-cosine phases.

To summarize, we have developed a mean-field theory of the CDW state
in magnetic field which takes into account imperfect nesting and the
shift of the CDW wave vector. This allows a detailed study of the
CDW properties and the phase diagram in high magnetic field below
the transition temperature. Our analysis gives the link between the
previous theoretical results based on the susceptibility calculation
\cite{ZBM} and the soliton solutions at zero
temperature.\cite{BDK,BrazKirovaReview} Although our study of the
CDW$_{2c}$ state is only applicable close to the transition line
$T_c(H)$ where the ration $\Delta_{\sigma}/T_c$ can be considered as
a small parameter, even an investigation in this region allows to
make some important conclusions about the structure of CDW at high
magnetic field. We have shown that the CDW$_x$ state at high
magnetic field has predominantly a double-cosine modulation with
wave vectors $\bfQ_{\pm }=\bfQ_{0}\pm q_x$. At perfect nesting the
one-cosine CDW with shifted wave vector exists (according to the
mean-field calculations) only in a very narrow region near the
tricritical point (see Fig. \ref{PhDTri} and Fig. \ref{PhDTot}). The
finite second (''antinesting'') harmonic in the electron dispersion
(\ref{dispersion}) does not change this picture considerably.
However, more specific dispersions or other perturbations may
substantially change the phase diagram. We also suggest an
interesting peculiarity of the cosine phase --- its spin asymmetry
that allows to make a controllable spin current.

P.G. is grateful to L.P. Gor'kov and M.V. Kartsovnik for many useful
discussions and to A. Melikidze for critical reading. The work was
supported by (PG) NSF Cooperative agreement No. DMR-0084173 and the
State of Florida, and, in part, by INTAS No 01-0791 and RFBR.


\begin{thebibliography}{99}
\bibitem{Gruner}  G. Gr\" uner, \emph{Density waves in Solids}
Perseus Publishing; 1st edition (January 15, 2000).

\bibitem{Monceau}  P. Monceau, \emph{Electronic Properties of
Inorganic Quasi One-Dimensional Compounds}, D. Reidel Pub. Co.;
(April 2002).

\bibitem{1Dtheory}  See e.g., the books: E. Fradkin,
\emph{Field Theories of Condensed Matter Systems}, Westview Press
(March, 1998); Alexander O. Gogolin, Alexander A. Nersesyan, Alexei
M. Tsvelik, \emph{Bosonization Approach to Strongly Correlated
Systems}, Cambridge University Press (December 10, 1998).

\bibitem{Comment1D} One illustrating example that the strictly 1D
models are not applicable to describe actual Q1D compounds could be
the phase diagram of low-temprature ground state in coordinates of
coupling constants. The renorm-group solution of the 1D models with
$g_{1,2}$ coupling constants ($g_{1}$ and $g_2$ being the backward
and forward scattering amplitudes respectively) (see, e.g., D.
Senechal, cond-mat/9908262) gives the gapless Luttinger-liquid state
at $|g_1|<2g_2$ while in actual compounds no such a state is usually
observed.

\bibitem{HorGutWeger} B. Horovitz, H. Gutfreund and M. Weger,
Phys Rev B. \textbf{12}, 3174 (1975)


\bibitem{McKenzieFluct} Ross H. McKenzie, Phys Rev B. \textbf{52},
16428 (1995)

\bibitem{Ishi}  T.~Ishiguro, K.~Yamaji, and G.~Saito, \emph{Organic
Superconductors}, 2$^{nd}$ edition, Springer-Verlag Berlin
Heidelberg, 1998.

\bibitem{FFLOFF} P. Fulde and A. Ferrel, Phys. Rev. 135, A550 (1964).

\bibitem{FFLOLO} A. I. Larkin and Yu. N. Ovchinnikov,
Sov. Phys. JETP 20, 762 (1965).

\bibitem{Su} W.P. Su, J. R. Schrieffer and A. J. Heeger,
Phys. Rev. Lett. \textbf{42}, 1698 (1979); Phys. Rev. B \textbf{22},
2099 (1980).

\bibitem{Braz} S.A. Brazovskij, Sov. Phys. JETP \textbf{51}, 342 (1980).

\bibitem{BGS} S.A. Brazovskii, L.P. Gor'kov, J.R. Schrieffer,
Physica Scripta \textbf{25}, 423 (1982).

\bibitem{BGL} S.A. Brazovskii, L.P. Gor'kov, A.G. Lebed',
Sov. Phys. JETP \textbf{56}, 683 (1982) [Zh. Eksp. Teor. Fiz.
\textbf{83}, 1198 (1982)].

\bibitem{BuzTug} A.I. Buzdin and V.V. Tugushev, Sov. Phys. JETP {\bf
58}, 428 (1983) [Zh. Eksp. Teor. Fiz. {\bf 85}, 735 (1983)].

\bibitem{SuReview} W.P. Su and J. R. Schrieffer, \emph{Physics in One Dimension}/
Ed. by J. Bernascony and T. Schneider, Springer series in Solid
State Sciences, Berlin, Heidelberg and New York: Springer, 1981.

\bibitem{BrazKirovaReview} S.A. Brazovskij and N.N. Kirova, Sov. Sci. Rev. A Phys.,
volume \textbf{5}, p. 99 (1984).

\bibitem{BDK} S.A. Brazovskii, I.E. Dzjaloshinskii and N.N. Kirova, Sov. Phys. JETP
\textbf{54}, 1209 (1981) [ZhETF 81, 2279 (1981)].

\bibitem{gor84}  L.~P.~Gor'kov and A.~G.~Lebed, J. Phys. (Paris) Lett.
\textbf{45}, L433 (1984).

\bibitem{Mont}  G.~Montambaux, M.~H\'{e}ritier, and P.~Lederer, Phys. Rev.
Lett. \textbf{55}, 2078 (1985).

\bibitem{FISDWEx} J.~F.~Kwak, J.~E.~P.M. Chaikin et al., Phys. Rev. Lett. {\bf 56},
972 (1986).

\bibitem{LebFICDW} A. G. Lebed, JETP Lett. {\bf 78}, 138 (2003).

\bibitem{DietrichFulde}  W.
Dieterich and P.Fulde, Z. Phys. \textbf{265}, 239 (1973).

\bibitem{cha96}  P.~M.~Chaikin, J. Phys. I France \textbf{6}, 1875 (1996).

\bibitem{ZBM}  D.~Zanchi, A.~Bjeli\v{s}, and G.~Montambaux, Phys. Rev. B
\textbf{53}, 1240 (1996).

\bibitem{Machida}  M. Fujita, K. Machida and H. Nakanishi, J. Phys. Soc Jpn., 54, 3820 (1985).

\bibitem{KartsLaukin} M.V. Kartsovnik and V.N. Laukin, J. Phys. I France {\bf 6},
1753 (1996).

\bibitem{Biskup} N.~Biskup, J.A.A.J.~Perenboom, J.S.~Qualls, and J.S.~Brooks,
Solid State Commun. 107, 503  (1998).

\bibitem{Christ} P.~Christ, W.~Biberacher, M.~V.~Kartsovnik, E.~Steep, E.~Balthes,
H.~Weiss, and H.~M\"uller, JETP Lett. 71, 303 (2000) [Pis'ma Zh.
Eksp. Teor. Fiz. 71, 437 (2000)].

\bibitem{Qualls} J.~S. Qualls, L. Balicas, J. S. Brooks et al.,
Phys. Rev. B {\bf 62}, 10008 (2000).

\bibitem{Andres} D. Andres, M. V. Kartsovnik, W. Biberacher et al.,
Phys. Rev. B 64, 161104(R) (2001).

\bibitem{HarrCDW} N. Harrison, C. H. Mielke, A. D. Christianson,
J.S. Brooks, and M. Tokumoto, Phys. Rev. Lett. 86, 1586 (2001).

\bibitem{OrbQuant} D. Andres, M. V. Kartsovnik, P. D. Grigoriev, W. Biberacher, and
H. M\"uller, Phys. Rev. B 68, 201101(R) (2003).

\bibitem{UncCDW1} K. Maki, B. D\"ora, M. V. Kartsovnik et al.,
Phys. Rev. Lett. 90, 256402 (2003).

\bibitem{UncCDW2} N. Harrison, J. Singleton, A. Bangura et al.,
Phys. Rev. B 69, 165103 (2004).

\bibitem{Graf} D. Graf, J. S. Brooks, E. S. Choi et al.,
Phys. Rev. B {\bf 69}, 125113 (2004).

\bibitem{Graf1} D. Graf, E. S. Choi, J. S. Brooks et al.,
Phys. Rev. Lett. {\bf 93}, 076406 (2004).

\bibitem{Per1} E. B. Lopes, M. J. Matos, R. T. Henriques, M. Almeida, and J. Dumas
Phys. Rev. B 52, R2237 (1995).

\bibitem{McKenzieCondMat} R.McKenzie, cond-mat/9706235 (1998)

\bibitem{BZ99} A.~Bjeli\v s, D.~Zanchi, and G.~Montambaux, ``Pauli and
orbital effects of magnetic field on charge density waves'', {\tt
cond-mat/9909303}.

\bibitem{HasFuk} Y.Hasegawa and H. Fukuyama, J. Phys. Soc. Japan \textbf{55}, 3978 (1986).

\bibitem{Dimitrova} O. V. Dimitrova and M. V. Feigel'man, JETP Lett. 78, 637 (2003)

\bibitem{BurkRainer} H. Burkhardt and D. Rainer, Ann. Physik
\textbf{3}, 181 (1994).

\end{thebibliography}
\end{document}